\documentclass[twocolumn, prl,showpacs,superscriptaddress]{revtex4}

\usepackage{mathrsfs} 
\usepackage{amssymb} 
\usepackage{graphicx}
\usepackage{subfigure}
\usepackage{dcolumn}
\usepackage{bm}

\begin{document}

\title{Optical Phonon Anomaly in Bilayer Graphene with Ultrahigh Carrier Densities}
\date{\today}

\author{Jia-An Yan}
\email{jiaanyan@gmail.com}
\affiliation{Department of Physics, Astronomy, and Geosciences, Towson University,\\
8000 York Road, Towson, MD 21252, USA}
\author{K. Varga}
\affiliation{Department of Physics and Astronomy, Vanderbilt University,
Nashville, TN 37035, USA}
\author{M. Y. Chou}
\email{meiyin.chou@physics.gatech.edu}
\affiliation{School of Physics, Georgia Institute of Technology,
Atlanta, Georgia 30332, USA}
\affiliation{Institute of Atomic and Molecular Sciences, Academia Sinica, Taipei 10617, Taiwan}

\begin{abstract}
Electron-phonon coupling (EPC) in bilayer graphene (BLG) at different doping levels is studied by first-principles calculations. The phonons considered are long-wavelength high-energy symmetric (S) and antisymmetric (AS) optical modes. Both are shown to have distinct EPC-induced phonon linewidths and frequency shifts as a function of the Fermi level $E_F$. We find that the AS mode has a strong coupling with the lowest two conduction bands when the Fermi level $E_F$ is nearly 0.5 eV above the neutrality point, giving rise to a giant linewidth (more than 100 cm$^{-1}$) and a significant frequency softening ($\sim$ 60 cm$^{-1}$). Our \emph{ab initio} calculations show that the origin of the dramatic change arises from the unusual band structure in BLG. The results highlight the band structure effects on the EPC in BLG in the high carrier density regime.
\end{abstract}


%

\maketitle

Electron-phonon coupling (EPC) is an important effect in monolayer graphene (MLG) \cite{Neto2009}. Interesting phenomena such as the renormalization of the phonon energy \cite{Ando2006}, the Kohn anomalies \cite{Piscanec2004}, and the breakdown of the adiabatic (Born-Oppenheimer) approximation \cite{Lazzeri2006,Pisana2007} have been reported. It is expected that even more intriguing effects should be found in AB-stacked bilayer graphene (BLG) where both the band structure and the doping level (i.e., the Fermi level $E_F$) can be tuned through the applied electrical gates \cite{Zhang2008,Li2009,Zhang2009,Mak2009}, allowing for the control of a delicate interplay between electrons, phonons, and photons \cite{Marlard2008,Ando2007,Ando2011,Yan2008,Kuzmenko2009,Tang2010}. Previous investigations mainly focused on situations with charge carriers near the charge neutrality Dirac point $E_D$ (i.e., $|E_F-E_D|<$0.4 eV) \cite{Marlard2008,Yan2008,Kuzmenko2009,Tang2010}. In this energy range, phenomena such as the phonon mode renormalization, Raman broadening, and the phonon frequency shift induced by EPC have been successfully predicted by a simplified tight-binding (TB) model and effective mass theory \cite{Ando2007,Ando2011}.

Recent progress in fabricating electrolytic \cite{Efetov2010,Efetov2011,Chen2011} and ionic-liquid gates \cite{Ye2010} provides the possibilities of doping MLG and BLG with ultrahigh charge-carrier densities of $|n| > 10^{14}$ cm$^{-2}$ and of tuning the Fermi level close to the van Hove singularity (VHS) point at $M$ in the first Brillouin zone (BZ) \cite{Efetov2010, McChesney2010}. Such a high carrier density will embark interesting technological applications including supercapacitors \cite{Stoller2008}, transparent electrodes \cite{Lee2010}, and high performance organic thin film transistors \cite{Cho2008}. In this regard, understanding the carrier dynamics will be a crucial step for the potential electronic device applications. Distinct many-body effects and superconducting instability have been observed in doped graphene when the Fermi energy approaches the VHS point \cite{McChesney2010}. Since superconductivity also occurs in graphene-related systems such as graphite-intercalation compounds (GICs), the study of EPC in BLG might shed new light on the underlying mechanism of superconductivity in GICs \cite{Valla2009,Gruneis2009,Zhao2011}.

Raman and infrared (IR) spectra are powerful tools to probe the EPC in MLG and BLG by providing the information of the phonon linewidth and frequency shifts as a function of doping and field strengths \cite{Lazzeri2006,Pisana2007,Yan2008,Kuzmenko2009,Das2009}. Although the aforementioned models \cite{Ando2007,Ando2011} are known to work well for low-energy charge carriers, deviations between theoretical predictions and experimental observations become significant when $|E_F-E_D|$ $>$ 0.4 eV \cite{Das2009}. Such a discrepancy calls for a full consideration of band structure effects from first principles, as the electronic structure in this regime is expected to be considerably modified from its low energy part.

\begin{figure}[tbp]
\centering
  \includegraphics[width=8.5cm]{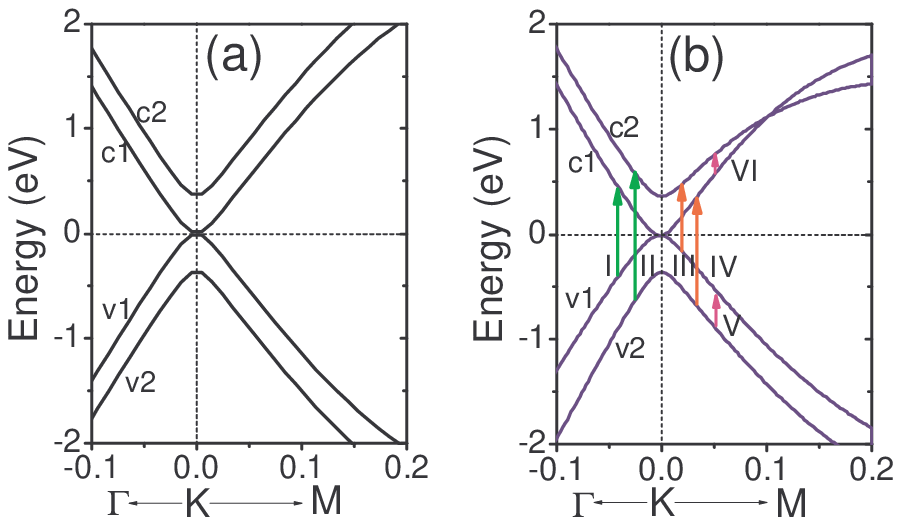}
 \caption{(Color online) Band structures of bilayer graphene calculated from (a) a simplified tight-binding model with only nearest-neighbor (NN) coupling and (b) density functional theory. The $x$ axis is in units of 2$\pi$/$a_0$, where $a_0$ is the graphene lattice constant. All possible transitions (I-VI) through the phonon modes are labeled in (b). The neutrality point $E_D$ has been shifted to energy zero.}\label{fig:band}
\end{figure}

In this work, we show that the band structure of BLG at high doping levels plays a critical role in the EPC of the two long-wavelength high-energy  symmetric (S) and antisymmetric (AS) optical modes in the high carrier density regime. For both modes, the linewidth $\gamma$ exhibits a dramatic dependence on the electron and hole doping level. In particular, the AS ($E_u$) mode exhibits a giant EPC-induced phonon linewidth (more than 100 cm$^{-1}$) and significant softening (60 cm$^{-1}$) when the Fermi level $E_F$ is 0.5 eV above the neutrality point $E_D$. Despite an intensive study of the EPC in BLG \cite{Park2008,Ando2007,Ando2011}, this feature has not been reported so far. This giant enhancement originates from the strong coupling between the AS mode and two conduction bands. The strong EPC in this high carrier-density regime might affect the performance of BLG-based devices.

The first-principles calculations reported in this work are performed using the Quantum ESPRESSO code \cite{Baroni2001, pwscf}. The electronic structure is obtained using the local density approximation (LDA) within the density-functional theory (DFT), and the core-valence interaction is modeled by norm-conserving pseudopotentials \cite{Troullier1991}. The wave functions of the valence electrons are expanded in plane waves with a kinetic energy cutoff of 70 Ry. A vacuum region of 20 \AA~ has been introduced
to eliminate the artificial interaction between neighboring supercells along the $z$ direction. The relaxed
C-C bond length is 1.42 \AA~  and the interlayer distance is 3.32 \AA~ for BLG.

The phonon frequencies and associated eigenvectors were
computed using the density-functional perturbation theory (DFPT)
\cite{Baroni2001}, details of which have been presented in our
previous work \cite{me2008,me2009}. The self-energy $\Pi_{\mathbf{q}\nu} (\omega)$ of a phonon with wave vector $\mathbf{q}$, branch index $\nu$, and frequency $\omega_{\mathbf{q}\nu}$ provides information on the renormalization and damping of that phonon due to the interaction with other elementary excitations. Following the Migdal approximation, the self-energy induced by the EPC in BLG reads \cite{Ando2007}:
\begin{widetext}
\begin{equation}
\Pi_{\mathbf{q}\nu}(\omega)
=
2 \sum_{mn}\int\frac{d\mathbf{k}}{\Omega_{\mathrm{BZ}}}|g_{mn}^{\nu}(\mathbf{k},\mathbf{q})|^2 \frac{[f(\epsilon_{n\mathbf{k+q}})-f(\epsilon_{m\mathbf{k}})][\epsilon_{n\mathbf{k+q}}-\epsilon_{m\mathbf{k}}]}{(\epsilon_{n\mathbf{k+q}}-\epsilon_{m\mathbf{k}})^2-(\hbar\omega+i\eta)^2},
\end{equation}
\end{widetext}
where $\epsilon_{m\mathbf{k}}$ is the energy of an electronic state $|m\mathbf{k}\rangle$ with crystal momentum $\mathbf{k}$ and band index $m$, $f(\epsilon_{m\mathbf{k}})$ the corresponding Fermi occupation, and $\eta$ a positive infinitesimal. For a given mode $\omega=\omega_0$, the phonon linewidth is $\gamma$ = $-2\mathrm{Im}(\Pi_{\mathbf{q}\nu} (\omega_0))$ and the phonon frequency shift is $\Delta \omega$ = $\frac{1}{\hbar}[\mathrm{Re}(\Pi_{\mathbf{q}\nu} (\omega_0)|_{E_F}-\Pi_{\mathbf{q}\nu}(\omega_0)|_{E_F=0}]$.

The EPC matrix element in Eq.~(1) is given by
\begin{eqnarray}
g_{mn}^{\nu}(\mathbf{k},\mathbf{q})=\sqrt{\frac{\hbar}{2M\omega_{\mathbf{q}}^{\nu}}}\langle
m\;\mathbf{k+q}|\frac{\delta V_{scf}}{\delta
u_{\mathbf{q}}^{\nu}}|n\;\mathbf{k}\rangle,
\end{eqnarray}
where $\delta V_{scf}\equiv
V_{scf}(u_{\mathbf{q}}^{\nu})-V_{scf}(0)$ is the variation of the
self-consistent potential field due to the perturbation of a phonon
with wave vector $\mathbf{q}$ and branch index $\nu$. Variations of the potential field $\delta V_{scf}$ are calculated through
self-consistent calculations to find the potential field for both
perturbed and unperturbed systems. The perturbed phonon mode is handled with the frozen-phonon approach \cite{me2009}. The DFT calculations have been carried out on a dense 201$\times$201 $k$-grid within a minizone (0.4$\times$0.4) enclosing the BZ corner $K$ in the reciprocal space. This is equivalent to 500$\times$500 $k$-grid sampling in the whole Brillouin zone. Finally, the EPC matrix elements are computed using Eq.~(2).
By changing the Fermi level $E_F$ in Eq.~(1), we can investigate the dependence of $\gamma$ and $\Delta \omega$ on different doping levels, assuming the EPC matrix elements are unchanged. This approximation is justified by the small dependence of the EPC matrix elements on doping for the $\Gamma$ phonon modes in graphene \cite{Attaccalite2010}. For all the linewidths calculated below, we used a parameter $\eta$ = 5 meV.

In Fig.~\ref{fig:band} we show the band structure of BLG obtained by the TB model (with only nearest-neighbor (NN) interactions) and DFT calculations. For BLG, the low-energy dispersions within a TB model can be well described by $E(k)$ = $\pm t_2/2\pm\sqrt{t_2^2/4+t_1^2|f(k)|^2}$, with $t_1$ = $-2.7$ eV and $t_2$ = 0.36 eV, the intralayer and interlayer hopping parameters, respectively \cite{Yan2011}. The low-energy parabolic dispersions predicted by the TB model are in agreement with the DFT result. Nevertheless, there is still a noticeable difference: In Fig.~\ref{fig:band}(b), the band dispersions have an evident asymmetry between the conduction and valence bands, while the TB model shows nearly symmetric band structure relative to the neutrality point [Fig.~\ref{fig:band}(a)]. This electron-hole asymmetry for the low-energy charge carriers has already been revealed by infrared spectroscopy \cite{Li2009} and becomes more significant for the high-energy carriers. In particular, our first-principles calculations predict that there is a crossing between the two conduction bands (c1, c2) at about 0.1$\times$$2\pi/a_0$ from $K$ along the $K$-$M$ direction in the BZ \cite{Yang2009}. The energy is around 1.0 eV above the neutrality point, as shown in Fig.~\ref{fig:band}(b). This feature is not captured by the simple TB model, indicating the necessity of first-principles calculations. As we will show later, the band structure plays a crucial role in understanding the physics of  high-energy charge carriers.

\begin{figure}[tbp]
\centering
   \includegraphics[width=7cm]{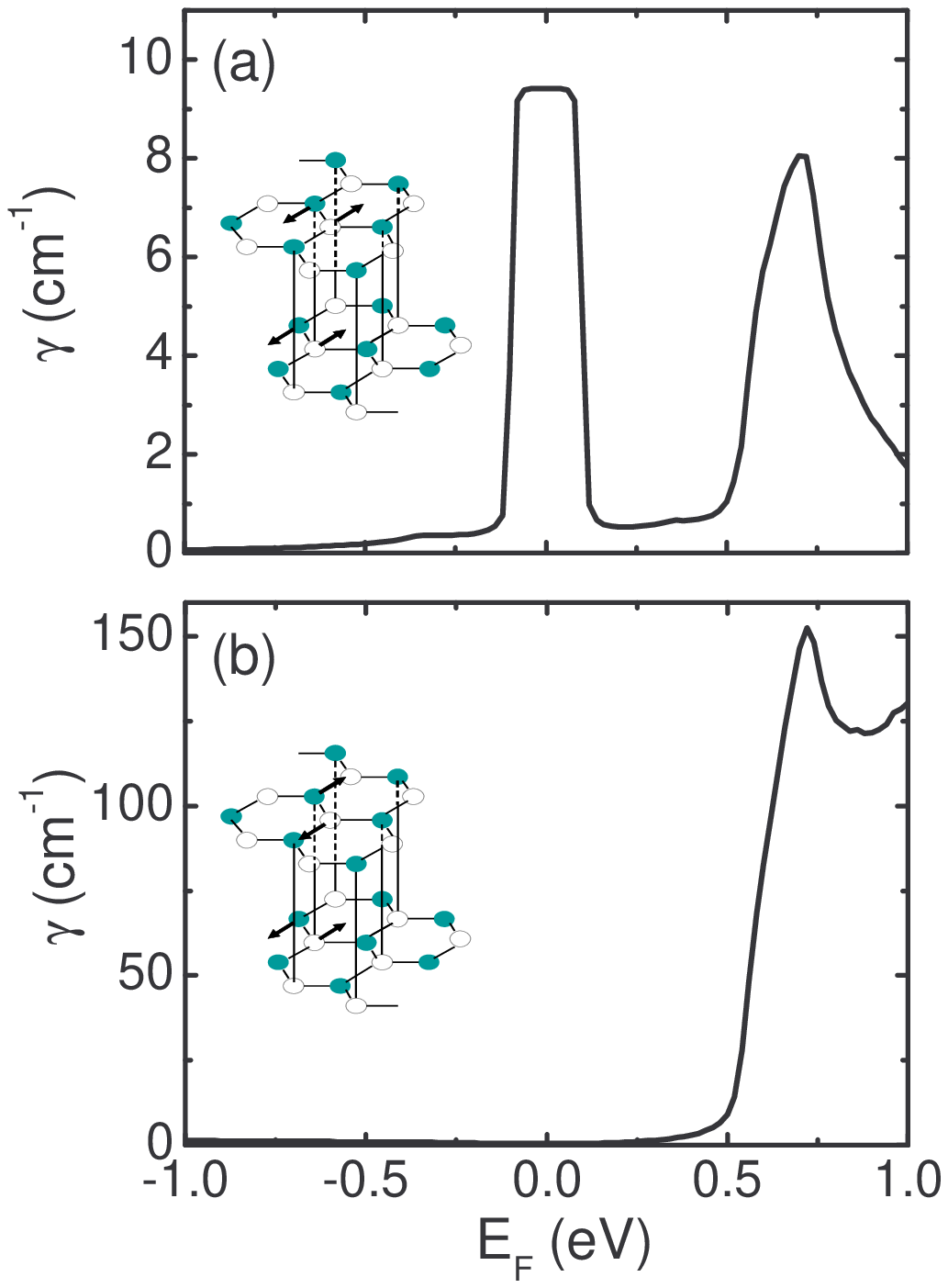}
 \caption{(Color online) Calculated linewidths $\gamma$ for the (a) symmetric (S) and (b) antisymmetric (AS) mode as a function of $E_F$. Note that the vertical scale has been changed in (b). The neutrality point $E_D$ is set to be energy zero. Insets are schematic plots of the S and AS modes. }\label{fig:gamma}
\end{figure}

The asymmetric band structure of BLG has important implications on the phonon linewidth and frequency shift because the S and AS modes couple with different bands in BLG \cite{me2009}. As indicated in Fig.~\ref{fig:band}(b), the allowed transitions [i.e., nonzero EPC matrix elements in Eq.~(2)] for the S mode are transitions I (v1-c1), II (v2-c2), V (v1-v2), and VI (c1-c2). In contrast, the AS mode only couples with transitions III (v1-c2), IV (v2-c1), V (v1-v2), and VI (c1-c2). When $E_F$ is tuned to be 0.5 eV above the neutrality point, the transition VI becomes active because the energy requirement is satisfied. On the other hand, the crossing of the two conduction bands (c1, c2) in this energy range provides a larger electronic phase space in the Brillouin zone for the phonon mode to couple with, as will be shown later. As a result, the linewidth of the AS mode is expected to be significantly enhanced.

\begin{figure}[tbp]
\centering
  \includegraphics[width=7cm]{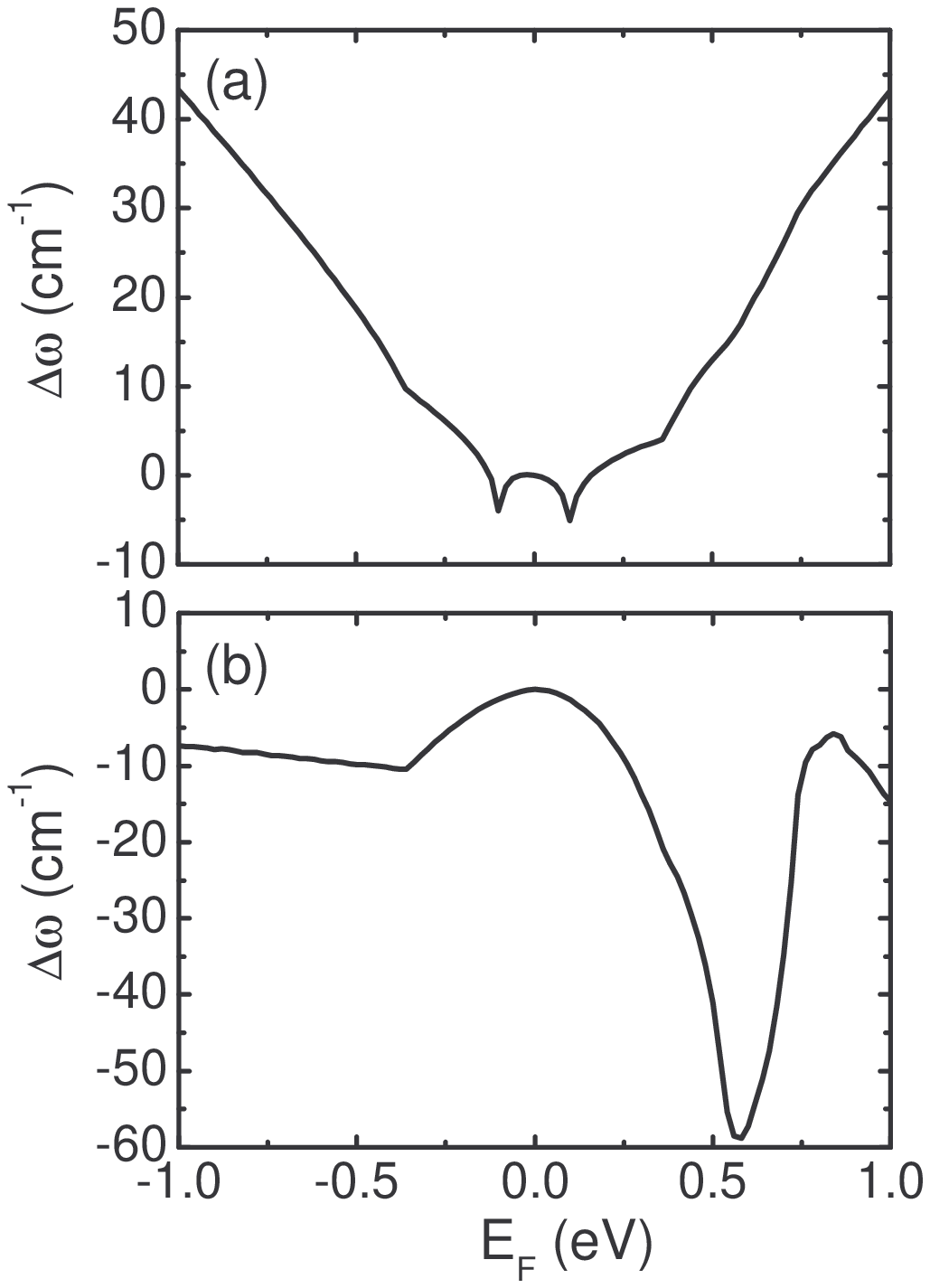}
 \caption{Calculated frequency shift $\Delta \omega$ for the (a) symmetric and (b) antisymmetric mode as a function of $E_F$. The neutrality point $E_D$ is set to be energy zero.}\label{fig:shift}
\end{figure}

In Figs.~\ref{fig:gamma}(a) and~\ref{fig:gamma}(b), we show the calculated linewidths for the S and AS modes with respect to $E_F$. The S and AS  modes ($\hbar\omega_0\sim$0.2 eV) considered in this work have been schematically shown in the insets. The corresponding frequency shifts $\Delta \omega$ are presented in Fig.~\ref{fig:shift}(a) and~\ref{fig:shift}(b), respectively.

In the low doping regime with $|E_F-E_D| < \hbar \omega_0/2 \sim 0.1$ eV, the S mode can be in resonant coupling with the electron-hole pair from the the top valence band (v1) and the bottom conduction band (c1). As a result, the linewidth of the S mode is constant, as depicted in Fig.~\ref{fig:gamma}(a). In contrast, the AS mode has a negligible linewidth in this range. Our calculations of phonon linewidth and frequency shift in this energy region reproduce the features found in previous DFT calculations \cite{Park2008} and agree reasonably well with the experimental data \cite{Yan2008}. This is also consistent with a previous study based on a continuum model \cite{Ando2007}.

In comparison, dramatic differences are found for the ultrahigh electron doping when $E_F -E_D > 0.5$ eV. The calculated $\gamma$ for the AS mode is significantly larger than that of the S mode. For example, when $E_F-E_D$ = 0.70 eV, $\gamma$ is 150 cm$^{-1}$ for the AS mode, around 20 times larger than that of the S mode (8.0 cm$^{-1}$). Correspondingly, there is a significant softening ($\sim 60$ cm$^{-1}$) for the frequency shift when $E_F = $0.6 eV, as shown in Fig.~\ref{fig:shift}(b). Further increase of the Fermi level will result in an even more significant increase of the phonon linewidth (not shown). In contrast, the phonon softening becomes smeared out as $E_F$ increases. Although such a doping level requires large electron densities, it may be achieved either by directly applying electrical gate in experiment \cite{Efetov2011} or by chemical doping \cite{McChesney2010}. For $E_F-E_D$ = 0.5 eV, the desired electron density in BLG is estimated to be $n \sim 6\times$10$^{13}$ cm$^{-2}$, still an order of magnitude smaller than the doping limit achieved in monolayer graphene ($\sim$ 4$\times$10$^{14}$ cm$^{-2}$) \cite{Efetov2010}. Therefore, we expect that this high electron doping regime will be realized in experiment \cite{Efetov2011} and that the giant phonon linewidth as well as the significant phonon mode softening may be verified by Raman or infrared measurements.

\begin{figure}[tbp]
\centering
  \includegraphics[width=7cm]{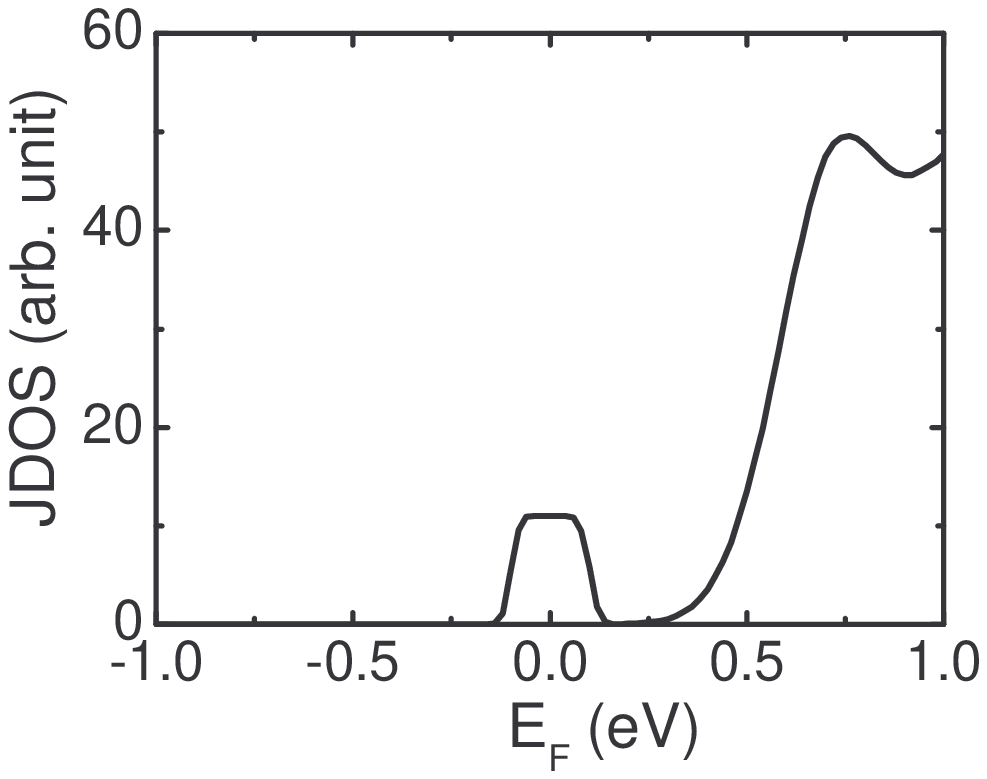}
 \caption{Calculated joint density of states (JDOS) $N(E_F,\omega_0)$ as a function of the Fermi level. The neutrality point $E_D$ is set to be energy zero.}\label{fig:jdos}
\end{figure}


This giant linewidth enhancement mainly arises from the large phase space associated with the high doping range. In Fig.~\ref{fig:jdos} we plot the joint density of states (JDOS) $N(E_F,\omega_0)=\frac{4\pi}{N_k}\sum_{\mathbf{k}jj'}[f(\epsilon_{\mathbf{k}j})-f(\epsilon_{\mathbf{k}j'})]\delta[\epsilon_{\mathbf{k}j}-\epsilon_{\mathbf{k}j'}+\hbar\omega_0]$ as a function of $E_F$. The JDOS increases dramatically when the Fermi level is in the range of $E_F-E_D >$ 0.5 eV, which allows more electronic states to couple with the phonon modes. The profile of $N(E_F)$ is similar to the phonon linewidth profile of the AS mode, indicating that the giant enhancement is mainly an electronic structure effect.

In summary, our first-principles calculations find that the phonon linewidths and frequency shifts for the long-wavelength high-energy optical modes in bilayer graphene exhibit a distinct dependence on the electron and hole doping due the intriguing interplay between the unique band structure and the phonon modes in this system. In particular, we predict that the linewidth for the antisymmetric mode could be significantly enhanced when the Fermi level is tuned to be 0.5 eV above the neutrality point.

J.A.Y. is grateful to Z. Jiang, F. Giustino, C. -H. Park, W. Duan, F. Liu and S. C. Zhang for fruitful discussions and thanks Mark A. Edwards for the support. Part of this work was done at the Georgia Southern University in Statesboro, Georgia. M.Y.C. acknowledges support by the US Department of Energy, Office of Basic Energy Sciences, Division of Materials Sciences and Engineering under Award No. DEFG02-97ER45632. This research used computational resources at the National Energy Research Scientific Computing Center (supported by the Office of Science of the U.S. Department of Energy under Contract No. DE-AC02-05CH11231) and at the National Institute for Computational Sciences under XSEDE startup allocation (Request No. DMR110111).


\begin{thebibliography}{99}

\bibitem{Neto2009} A. H. Castro Neto, F. Guinea, N. M. R. Peres, K. S. Novoselov and A. K. Geim, Rev. Mod. Phys. \textbf{81}, 109 (2009) and references therein.

\bibitem{Ando2006} T. Ando, J. Phys. Soc. Jpn. \textbf{75}, 124701 (2006).

\bibitem{Piscanec2004} S. Piscanec, M. Lazzeri, F. Mauri, A. C. Ferrari, and J.
Robertson, Phys. Rev. Lett. \textbf{93}, 185503 (2004).

\bibitem{Pisana2007}S. Pisana, M. Lazzeri, C. Casiraghi, K. S. Novoselov, A. K. Geim, A. C. Ferrari, and F. Mauri, Nat. Mater. \textbf{6}, 198 (2007).

\bibitem{Lazzeri2006} Michele Lazzeri and Francesco Mauri, Phys. Rev. Lett. \textbf{97}, 266407 (2006).

\bibitem{Zhang2008} L. M. Zhang, Z. Q. Li, D. N. Basov, and M. M. Fogler, Z. Hao and M. C. Martin, Phys. Rev. B \textbf{78}, 235408 (2008).

\bibitem{Li2009} Z. Q. Li, E. A. Henriksen, Z. Jiang, Z. Hao, M. C. Martin, P. Kim, H. L. Stormer, and D. N. Basov, Phys. Rev. Lett. \textbf{102}, 037403 (2009). 

\bibitem{Mak2009} K. F. Mak, C. H. Lui, J. Shan, and T. F. Heinz, Phys. Rev. Lett. \textbf{102}, 256405 (2009).

\bibitem{Zhang2009} Y. Zhang, T.-Ta Tang, C. Girit, Z. Hao, M. C. Martin, A. Zettl, M. F. Crommie, Y. R. Shen, and F. Wang,
Nature \textbf{459}, 820 (2009). 

\bibitem{Marlard2008} L. M. Malard, D. C. Elias, E. S. Alves, and M. A. Pimenta, Phys. Rev. Lett. \textbf{101}, 257401 (2008).

\bibitem{Yan2008} J. Yan, E. A. Henriksen, P. Kim, and A. Pinczuk, Phys. Rev. Lett. \textbf{101}, 136804 (2008).

\bibitem{Kuzmenko2009} A. B. Kuzmenko, L. Benfatto, E. Cappelluti, I. Crassee, D. van der Marel, P. Blake, K. S. Novoselov, and A. K. Geim, Phys. Rev. Lett. \textbf{103}, 116804 (2009).

\bibitem{Ando2007} T. Ando, J. Phys. Soc. Jpn. \textbf{76}, 104711 (2007).


\bibitem{Ando2011} T. Ando, Physica E \textbf{43}, 645 (2011).


\bibitem{Tang2010} T.-T. Tang, Y. Zhang, C.-H. Park, B. Geng, C. Girit, Z. Hao, M. C. Martin, A. Zettl, M. F. Crommie, S. G. Louie, Y. R. Shen and F. Wang, Nat. Nanotech. \textbf{5}, 32 (2010).



\bibitem{Efetov2010} D. K. Efetov and P. Kim, Phys. Rev. Lett. \textbf{105}, 256805 (2010).


\bibitem{Efetov2011} D. K. Efetov, P. Maher, S. Glinskis, and P. Kim, Phys. Rev. B \textbf{84}, 161412(R) (2011).


\bibitem{Chen2011} Chi-Fan Chen, Cheol-Hwan Park, Bryan W. Boudouris, Jason Horng, Baisong Geng, Caglar Girit, Alex Zettl, Michael F. Crommie, Rachel A. Segalman, Steven G. Louie, and Feng Wang, Nature \textbf{471}, 617 (2011).


\bibitem{Ye2010} J. T. Ye, M. F. Craciun, M. Koshino, S. Russo, S. Inoue, H. T. Yuan, H. Shimotani, A. F. Morpurgo, Y. Iwasa, Proc. Natl. Acad. Sci. USA \textbf{108}, 13002 (2011).


\bibitem{McChesney2010} J. L. McChesney, A. Bostwick, T. Ohta, T. Seyller, K. Horn, J. Gonzalez, and E. Rotenberg, Phys. Rev. Lett. \textbf{104}, 136803 (2010).


\bibitem{Stoller2008} M. D. Stoller, S. J. Park, Y. W. Zhu, J. H. An, and R. S. Ruoff, Nano Lett. \textbf{8}, 3498 (2008).

\bibitem{Lee2010} B. Lee, Y. Chen, F. Duerr, D. Mastrogiovanni, E. Garfunkel, E. Y. Andrei, and V. Podzorov, Nano Lett. \textbf{10}, 2427 (2010).

\bibitem{Cho2008} J. H. Cho, J. Lee, Y. He, B. Kim, T. P. Lodge, and C. D.
Frisbie, Adv. Mater. \textbf{20}, 686 (2008).

\bibitem{Valla2009} T. Valla, J. Camacho, Z.-H. Pan, A. V. Fedorov, A. C. Walters, C. A. Howard, and M. Ellerby,  Phys. Rev. Lett. \textbf{102}, 107007 (2009).


\bibitem{Gruneis2009} A. Gruneis, C. Attaccalite, A. Rubio, D. V. Vyalikh, S. L. Molodtsov,
J. Fink, R. Follath, W. Eberhardt, B. Buchner, T. Pichler, Phys.
Rev. B \textbf{79}, 205106 (2009).

\bibitem{Zhao2011} W. Zhao, P. H. Tan, J. Liu, and A. C. Ferrari, J. Am. Chem. Soc. \textbf{133}, 5941 (2011).






\bibitem{Das2009} A. Das, B. Chakraborty, S. Piscanec, S. Pisana, A. K. Sood, and A. C. Ferrari, Phys. Rev. B \textbf{79}, 155417 (2009).


\bibitem{Park2008} C.-H. Park, F. Giustino, M. L. Cohen, and S. G. Louie, Phys. Rev. Lett. \textbf{99}, 086804 (2007); Nano Lett. \textbf{8}, 4229 (2008).

\bibitem{Baroni2001} S. Baroni, S. de Gironcoli, and A. Dal Corso, Rev. Mod.
Phys. \textbf{73}, 515 (2001).


\bibitem{pwscf} P. Giannozzi, et al., J. Phys. Condens. Matter \textbf{21}, 395502 (2009).


\bibitem{Troullier1991}  N. Troullier and J. L. Martins, Phys. Rev. B \textbf{43}, 1993 (1991).

\bibitem{me2008} Jia-An Yan, W. Y. Ruan, and M. Y. Chou, Phys. Rev. B \textbf{77}, 125401 (2008)

\bibitem{me2009} Jia-An Yan, W. Y. Ruan, and M. Y. Chou, Phys. Rev. B \textbf{79}, 115443 (2009).








\bibitem{Attaccalite2010} C. Attaccalite, L. Wirtz, M. Lazzeri, F. Mauri, and A. Rubio, Nano Lett. \textbf{10}, 1172 (2010).
%


\bibitem{Yan2011} Jia-An Yan, W. Y. Ruan, and M. Y. Chou, Phys. Rev. B \textbf{83}, 245418 (2011).

\bibitem{Yang2009} The band dispersions of BLG with GW corrections exhibit a similar trend for the two conduction bands. See Li Yang, et al., Phys. Rev. Lett. \textbf{103}, 186802 (2009).

%
%
%
%


\end{thebibliography}
\end{document}